\documentclass{article}
\usepackage{waspaa21,amsmath,url,times}
\usepackage{color}
\usepackage{amsmath}
\usepackage{bm,amsfonts,here,subfig}
\usepackage{algorithm}
\usepackage{algpseudocode}
\usepackage{graphicx}

\usepackage{cite}

\newcommand{\argmin}{\mathop{\rm arg~min}\limits}

\newcommand{\minimize}{\mathop{\rm minimize}\limits}

\title{Mean-square-error-based secondary source placement\\in sound field synthesis with prior information on desired field}

\name{Keisuke Kimura, Shoichi Koyama, Natsuki Ueno, and Hiroshi Saruwatari}
\address{The University of Tokyo, 7-3-1 Hongo, Bunkyo-ku, Tokyo 113-8656, Japan}

\begin{document}

\ninept
\maketitle

\begin{sloppy}

\begin{abstract}
A method of optimizing secondary source placement in sound field synthesis is proposed. Such an optimization method will be useful when the allowable placement region and available number of loudspeakers are limited. We formulate a mean-square-error-based cost function, incorporating the statistical properties of possible desired sound fields, for general linear-least-squares-based sound field synthesis methods, including pressure matching and (weighted) mode matching, whereas most of the current methods are applicable only to the pressure-matching method. An efficient greedy algorithm for minimizing the proposed cost function is also derived. Numerical experiments indicated that a high reproduction accuracy can be achieved by the placement optimized by the proposed method compared with the empirically used regular placement.
\end{abstract}


\vspace{-6pt}
\section{Introduction}
\label{sec:intro}
\vspace{-6pt}

Sound field synthesis/reproduction methods are aimed at high-fidelity spatial audio for virtual/augmented reality, generating multiple sound zones for personal audio, cancelling incoming sound for noise reduction, and so forth. Here, we focus on a problem of synthesizing a (given) desired sound field inside a target region with multiple loudspeakers. 
There are two categories of sound field synthesis methods. One is analytical methods based on boundary integral equations analytically derived from the Helmholtz equation, such as wave field synthesis and higher-order ambisonics~\cite{Berkhout:JASA_J_1993,Spors:AES124conv,Poletti:J_AES_2005,Ahrens:Acustica2008,Wu:IEEE_J_ASLP2009,Koyama:IEEE_J_ASLP2013,Koyama:JASA_J_2016}. The other is numerical methods based on numerical optimization to minimize error between synthesized and desired sound fields inside the target region, such as pressure matching and mode matching~\cite{Miyoshi:IEEE_J_ASSP_1988,Kirkeby:JASA_J_1996,Daniel:AES114conv,Poletti:J_AES_2005,Gauthier:JASA_J_2005,Betlehem:JASA_J_2005,Ueno:IEEE_ACM_J_ASLP2019}. In the analytical methods, the loudspeakers are arranged on the surface of a simple shape, such as a sphere, plane, circle, and line, and the driving signals are obtained from a discrete approximation of the integral equation. On the other hand, in the numerical methods, the loudspeaker placement can be flexible, and the driving signals are generally derived as a closed-form least-squares solution. 

The flexibility of the loudspeaker placement in the numerical methods is useful in practical applications; however, the reproduction accuracy highly depends on the placement, especially when the allowable placement region and available number of loudspeakers are limited. If the target environment is anechoic, the possible placement region is on the surface of a simple shape, and the desired sound field can be sound waves from all directions, the optimal choice will be a regular placement on the surface as used in the analytical methods. However, if the loudspeaker placement can be optimized, depending on the target room environment and allowable placement region, a highly efficient sound field synthesis system with the smallest possible number of loudspeakers can be achieved. Moreover, when the possible desired sound fields are limited, the loudspeaker placement exploiting this limitation can be further efficient, for example, by reducing the loudspeakers on the directions that the sound waves are unlikely to be synthesized. 

Several attempts to optimize secondary source placement have been made~\cite{Snyder:JSV1991,Asano:IEEE_J_SAP_1999,Khalilian:IEEE_ACM_J_ASLP_2016,Koyama:ICASSP2018}, and they were summarized in a recent article~\cite{Koyama:IEEE_ACM_J_ASLP2020}. However, most of the methods are applicable only to the pressure-matching method, which is based on the discretization of the target region into control points. Moreover, properties of the desired sound fields are not well incorporated. For example, the empirical interpolation method~\cite{Koyama:ICASSP2018} is not applicable to the mode-matching methods, which is based on controlling the expansion coefficients of the sound field by using spherical/cylindrical wavefunctions~\cite{Daniel:AES114conv,Poletti:J_AES_2005,Ueno:IEEE_ACM_J_ASLP2019}, and cannot exploit prior knowledge on the desired sound field. In the methods based on sparse representation of the loudspeaker driving signals~\cite{Khalilian:IEEE_ACM_J_ASLP_2016,Lilis:5443604,routray:EUSIPCO2018,gauthier:JASA2017}, the loudspeakers are selected from the candidates for synthesizing a single specific desired sound field. 

In this study, we propose a method of optimizing the secondary source placement from predefined candidate points that is applicable to general sound field synthesis methods based on the linear-least-squares problem, which includes not only the pressure-matching method, but also the  mode-matching method and their variants. We formulate the cost function on the basis of the mean square error of the reproduced sound field, incorporating a statistical property on the possible desired sound fields, both for narrowband and broadband target frequencies. An efficient greedy algorithm for minimizing this cost function is developed. Numerical experiments in a two-dimensional (2D) sound field are conducted to evaluate the proposed method.

\vspace{-6pt}
\section{Problem statement}
\label{sec:2}
\vspace{-6pt}

Suppose that a target control region $\Omega$ is set in a space, and $L$ secondary sources, i.e., loudspeakers, are placed at the positions $\bm{r}_l$ ($l\in\{1,\ldots,L\}$). The driving signal of the $l$th loudspeaker at the angular frequency $\omega$ and its transfer function at the position $\bm{r}$ are denoted by $d_{l}(\omega)$ and $G(\bm{r}|\bm{r}_l,\omega)$, respectively. The synthesized pressure distribution at $\omega$ and $\bm{r}\in\Omega$, $u_{\mathrm{syn}}(\bm{r},\omega)$, is represented by the linear combination of $d_l$ and $G(\bm{r}|\bm{r}_l,\omega)$ as
\begin{align}
u_{\rm syn}(\bm{r},\omega)=\sum_{l=1}^L d_l(\omega) G(\bm{r}|\bm{r}_l,\omega) = \bm{g}(\bm{r},\omega)^{\mathsf{T}} \bm{d}(\omega), 
\label{eq:defsyn}
\end{align}
where $\bm{g}(\bm{r},\omega)\in\mathbb{C}^L$ and $\bm{d}(\omega)\in \mathbb{C}^L$ are the vectors consisting of $G(\bm{r}|\bm{r}_l,\omega)$ and $d_l(\omega)$, respectively. Hereafter, $\omega$ is omitted for notational simplicity. 

The aim of the sound field synthesis is to equalize the synthesized sound field $u_{\mathrm{syn}}(\bm{r})$ with the desired sound field, denoted by $u_{\mathrm{des}}(\bm{r})$, by controlling $\bm{d}$. Thus, the optimization problem for the sound field synthesis is described as the regional integral of the square reproduction error as
\begin{equation}
    \minimize_{\bm{d}\in\mathbb{C}^L}\ Q(\bm{d}) := \int_{\bm{r}\in\Omega} \left| \bm{g}(\bm{r})^{\mathsf{T}}\bm{d} - u_{\mathrm{des}}(\bm{r})\right|^2 \mathrm{d}\bm{r}.
    \label{eq:cost_synth}
\end{equation}
In the following, we consider a 2D sound field, but our formulations can be straightforwardly extended to a three-dimensional (3D) sound field. 

\begin{figure}[t]
  \centering
  \includegraphics[width=160pt,clip]{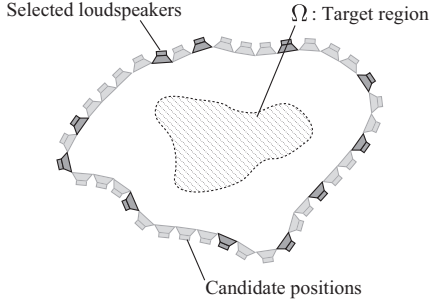}
  \vspace{-5pt}
  \caption{Sound field synthesis and its secondary source placement}
  \label{fig:sfs}
  \vspace{-10pt}
\end{figure}
\vspace{-4pt}
\subsection{Mode-matching-based sound field synthesis}
\vspace{-4pt}
The minimization problem of $Q(\bm{d})$ is difficult to solve owing to the regional integral. In the pressure-matching method, the target region $\Omega$ is discretized into control points, which leads to a simple linear-least-squares problem. Then, the driving signals are obtained as a closed-form least-squares solution minimizing the reproduction errors at the control points~\cite{Kirkeby:JASA_J_1996,Gauthier:JASA_J_2005}. 

The mode-matching method~\cite{Poletti:J_AES_2005} is a sound field synthesis method based on the control of expansion coefficients by cylindrical wavefunctions in 2D, defined in the polar coordinates $\bm{r}=(r,\phi)$ as
\begin{equation}
    \psi_m(\bm{r}) = J_m(kr) e^{\mathrm{j}m\phi},
\end{equation}
where $J_m(\cdot)$ is the $m$th-order Bessel function. One of its extensions is the weighted mode-matching method~\cite{Ueno:ICASSP2017,Ueno:IEEE_ACM_J_ASLP2019}, where the weight of each coefficient is analytically obtained on the basis of the regional integral of $\psi_m(\bm{r})$ to avoid the empirical truncation of the expansion order. The cost function of the weighted mode-matching is formulated as a sum of a proxy of $Q(\bm{d})$ and the regularization term of $\bm{d}$ as
\begin{align}
    F(\bm{d}) = \left( \bm{C}\bm{d} - \bm{b} \right)^{\mathsf{H}} \bm{W} \left( \bm{C}\bm{d} - \bm{b} \right) + \lambda \|\bm{d}\|^2,
    \label{eq:objfuncwmm}
\end{align}
where $\bm{C}\in\mathbb{C}^{(2M+1) \times L}$ and $\bm{b}\in\mathbb{C}^{2M+1}$ are the properly ordered matrix and vector of the expansion coefficients of $G(\bm{r}|\bm{r}_l)$ and $u_{\mathrm{des}}(\bm{r})$, respectively, $M$ is a sufficiently large truncation order,  $\bm{W}\in\mathbb{C}^{(2M+1)\times(2M+1)}$ is a Hermitian matrix of weighting factors, $\|\cdot\|$ denotes the $\ell_2$-norm, and $\lambda$ is the regularization parameter. By appropriately setting the weighting matrix $\bm{W}$, we obtain the first term of $F(\bm{d})$ that fairly approximates the original cost function $Q(\bm{d})$. The $(m,n)$th element of $\bm{W}$, $\bm{W}_{m,n}$, is calculated as
\begin{equation}
    \bm{W}_{m,n} = \int_{\bm{r}\in\Omega} \psi_m(\bm{r})^{\ast} \psi_n(\bm{r}) \mathrm{d}\bm{r}.
\end{equation}
Although a numerical integration is necessary to derive the elements of $\bm{W}$, a technique to avoid it exists for some simple shapes of $\Omega$~\cite{Ueno:ICASSP2017,Ueno:IEEE_ACM_J_ASLP2019}. The driving signal $\bm{d}$ minimizing $F(\bm{d})$ is thus obtained as
\begin{align}
\hat{\bm{d}} = (\bm{C}^{\mathsf{H}}\bm{W}\bm{C}+\lambda \bm{I})^{-1}\bm{C}^{\mathsf{H}}\bm{W}\bm{b}.
\label{eq:dhatwmm}
\end{align}
By setting $\bm{W}$ as the identity matrix, we obtain the driving signal $\hat{\bm{d}}$ of \eqref{eq:dhatwmm} that corresponds to that of the standard mode-matching method.

\vspace{-4pt}
\subsection{Secondary source placement problem}
\vspace{-4pt}
In the weighted mode-matching method as well as pressure-matching method, the loudspeakers can be arbitrarily placed in theory; however, their placements have a large effect on the reproduction accuracy, especially when the possible loudspeaker positions are restricted. The objective of this study is to develop an algorithm for optimizing the secondary source placement. 

Since optimizing the source placement in a continuous domain is impractical, we first set a discrete set of $N$ candidate positions of the loudspeakers $\mathcal{C}=\{\bm{r}_1, \ldots, \bm{r}_N\}\subset\mathbb{R}^2$ as shown in Fig.~\ref{fig:sfs}. We consider selecting the optimal subset of $L$ loudspeaker positions $\mathcal{S}=\{\bm{r}_1,\ldots,\bm{r}_L\} \subset \mathcal{C}$ under a certain measure. In the next section, we describe our proposed source placement method, using the weighted mode-matching method as an example, but our method can be applied to general linear-least-squares-based sound field synthesis methods, such as standard mode-matching and pressure-matching methods by appropriately replacing $\bm{C}$, $\bm{W}$, and $\bm{b}$.

\vspace{-6pt}
\section{Optimization of secondary source placement}
\label{sec:3}
\vspace{-6pt}

\subsection{Objective function based on mean square error}
\vspace{-4pt}
We formulate the objective function for the source placement based on the mean square error of the sound field synthesis. First, the $N$-length source selection vector is defined as $\bm{\varphi}^{(l)}\in\{0,1\}^N$, where only the element corresponding to the selected $l$th source, i.e., $l\in\mathcal{S}$, is $1$ and the others are $0$. The source selection matrix for $\mathcal{S}$ is also defined as $\bm{\Phi}_\mathcal{S}=[\bm{\varphi}^{(1)},\ldots,\bm{\varphi}^{(L)}]\in\{0,1\}^{N \times L}$. 

Now, we redefine $\bm{C}$ as the expansion coefficient matrix of the transfer functions of the candidate positions, i.e., $\bm{C}\in\mathbb{C}^{(2M+1) \times N}$, which can be given by using an analytical model or numerical simulation of $G(\bm{r}|\bm{r}_n)$ ($\bm{r}_n\in\mathcal{C}$). Thus, the expansion coefficient matrix of the transfer functions of the selected sources can be obtained as the product $\bm{C}\bm{\Phi}_{\mathcal{S}}$ ($\in\mathbb{C}^{(2M+1) \times L}$). The cost function of the weighted mode matching is reformulated by using $\bm{\Phi}_{\mathcal{S}}$ as
\begin{align}
F(\bm{d}) 
= (\bm{b} - \bm{C} \bm{\Phi}_{\mathcal{S}} \bm{d})^{\mathsf{H}} \bm{W}(\bm{b} - \bm{C} \bm{\Phi}_{\mathcal{S}} \bm{d}) 
+\lambda \| \bm{d} \|^2.
\label{eq:objfuncopt}
\end{align}
The driving signal minimizing \eqref{eq:objfuncopt} is given by
\begin{align}
  \hat{\bm{d}} = (\bm{\Phi}_{\mathcal{S}}^{\mathsf{H}} \bm{C}^{\mathsf{H}} \bm{W} \bm{C} \bm{\Phi}_{\mathcal{S}} + \lambda \bm{I})^{-1}\bm{\Phi}_{\mathcal{S}}^{\mathsf{H}} \bm{C}^{\mathsf{H}} \bm{W} \bm{b}. 
  \label{eq:dhatopt}
\end{align}

The cost function value of the weighted mode matching by using the estimated driving signal $\hat{\bm{d}}$ is derived by substituting \eqref{eq:dhatopt} into \eqref{eq:objfuncopt} as
\begin{align}
\!\!F(\hat{\bm{d}}) &= (\bm{b} - \bm{C} \bm{\Phi}_{\mathcal{S}} \hat{\bm{d}})^{\mathsf{H}} \bm{W} (\bm{b} - \bm{C} \bm{\Phi}_{\mathcal{S}} \hat{\bm{d}}) + \lambda  \hat{\bm{d}}^{\mathsf{H}}\hat{\bm{d}} \notag\\
&= \bm{b}^{\mathsf{H}} \left(\bm{W} -\bm{W}\bm{C}\bm{\Phi}_\mathcal{S}\bm{A}\bm{\Phi}_\mathcal{S}^\mathsf{H}\bm{C}^\mathsf{H}\bm{W}^\mathsf{H}\right) \bm{b}, 
\label{eq:quadratic}
\end{align}
where $\bm{A}= (\bm{\Phi}_{\mathcal{S}}^{\mathsf{H}} \bm{C}^{\mathsf{H}} \bm{W} \bm{C} \bm{\Phi}_{\mathcal{S}}+\lambda \bm{I})^{-1}$ is defined, and the fact that both $\bm{A}$ and $\bm{W}$ are Hermitian matrices is used for formulation. This quadratic form for $\bm{b}$ means the regional square error of the reproduced sound field when using the selected sources $\mathcal{S}$ and the desired expansion coefficients $\bm{b}$, including the regularization, because of the definition of the original cost function \eqref{eq:cost_synth}. We define the cost function for the source placement as the expectation value of $F(\hat{\bm{d}})$ with respect to the desired expansion coefficients $\bm{b}$ as
\begin{align}
J(\mathcal{S}) = \mathbb{E}[F(\hat{\bm{d})}] = \mathrm{trace}(\bm{D}\bm{\Sigma}) + \bm{\mu}^{\mathsf{H}}\bm{D}\bm{\mu}, 
\label{eq:costfunc}
\end{align}
where $\bm{\mu}$ and $\bm{\Sigma}$ are the mean vector and covariance matrix of $\bm{b}$, respectively, and 
\begin{align}
& \bm{D} =  \bm{W} -\bm{W}\bm{C}\bm{\Phi}_\mathcal{S}\bm{A}\bm{\Phi}_\mathcal{S}^\mathsf{H}\bm{C}^\mathsf{H}\bm{W}^\mathsf{H}. 
\label{eq:D}
\end{align}
Given $\bm{\mu}$ and $\bm{\Sigma}$ by using prior knowledge on the possible desired sound fields, the mean-square-error-based cost function for the source selection can be obtained. For example, $\bm{\mu}$ and $\bm{\Sigma}$ can be analytically calculated by integrating possible range of planewave directions of the desired fields. Thus, the problem to be solved is finding the subset $\mathcal{S}$ that minimizes $J(\mathcal{S})$ from $\mathcal{C}$.
\vspace{-4pt}
\subsection{Greedy algorithm of secondary source placement and its efficient computation}
\vspace{-4pt}
Since the minimization problem of $J(\mathcal{S})$ is a combinatorial problem, it is difficult to find the global optimal solution within a practical execution time. In particular, since the candidate source positions should be set at sufficiently small intervals compared with the target wavelength, the size of $\mathcal{C}$, $|\mathcal{C}|$, can be very large, and an exhaustive search of the optimal $\mathcal{S}$ is impractical. 

To obtain the approximate solution for minimizing $J(\mathcal{S})$, we propose a greedy algorithm for the source placement. As in Algorithm~\ref{alg:greedy}, the source position minimizing $J(\mathcal{S})$ is selected one by one. The stopping condition can be defined as the predefined maximum number of sources ($L$) or a threshold for the decrease in the cost function value at each iteration. 

\begin{algorithm}[t]
  \caption{Proposed algorithm for secondary source placement}  \label{alg:greedy}               
  \begin{algorithmic}
  \Require A set of candidate positions $\mathcal{C}$, and mean $\bm{\mu}$ and covariance $\bm{\Sigma}$ of $\bm{b}$
  \Ensure Secondary source selection $\mathcal{S}$
\State Initialize $\mathcal{S}=\emptyset$, $\bm{\Phi}_{\mathcal{S}}$, and $l=1$
\While{the stopping condition is not satisfied}
  \State Find $\hat{\bm{r}}_l = \argmin_{\bm{r}_l \in \mathcal{C}\backslash\mathcal{S}} J(\mathcal{S} \cup \{\bm{r}_{l}\})$ 
  \State $\mathcal{S}\leftarrow\mathcal{S} \cup \{ \hat{\bm{r}}_l \} $ and update $\bm{\Phi}_{\mathcal{S}}$
  \State $l \leftarrow l+1$
    \EndWhile
  \end{algorithmic}
\end{algorithm}

In Algorithm~\ref{alg:greedy}, it is necessary to compute the cost function $J(\mathcal{S})$ for all the unselected candidates at each iteration. By calculating the matrices $\bm{C}^{\mathsf{H}}\bm{W}\bm{C}$ and $\bm{W}\bm{C}$ in advance, we can obtain their products with $\bm{\Phi}_{\mathcal{S}}$ by extracting the corresponding submatrices from them; however, the $|\mathcal{C}\backslash\mathcal{S}|$ times computation of the inverse matrix $\bm{A}$ at each iteration is still costly. When the selected number of sources is $L$, the total computational cost of Algorithm~\ref{alg:greedy} regarding the computation of $\bm{A}$ can be estimated as $O(NL^4)$. This repeated computation can be avoided by using an alternative expression. Here, we denote the set of the source selection and matrix $\bm{A}$ at the $l$th iteration as $\mathcal{S}^{(l)}=\{\bm{r}_1,\ldots,\bm{r}_l\}$ and $\bm{A}^{(l)}$, respectively. At the $(l+1)$th iteration, the matrix $\bm{A}^{(l+1)}$ can be represented by using that at the previous iteration $\bm{A}^{(l)}$ as
\begin{align}
    &\bm{A}^{(l+1)} \notag\\
    &= \left( \bm{\Phi}_{\mathcal{S}^{(l+1)}}^{\mathsf{H}} \bm{C}^{\mathsf{H}} \bm{W} \bm{C} \bm{\Phi}_{\mathcal{S}^{(l+1)}} + \lambda \bm{I} \right)^{-1} \notag\\
    &=
    \begin{bmatrix}
    (\bm{A}^{(l)})^{-1} & \bm{\Phi}_{\mathcal{S}^{(l)}}^{\mathsf{H}} \bm{C}^{\mathsf{H}} \bm{W} \bm{C} \bm{\varphi}^{(l+1)}\\
    (\bm{\varphi}^{(l+1)})^{\mathsf{H}} \bm{C}^{\mathsf{H}} \bm{W} \bm{C} \bm{\Phi}_{\mathcal{S}^{(l)}} & (\bm{\varphi}^{(l+1)})^{\mathsf{H}} \bm{C}^{\mathsf{H}} \bm{W} \bm{C} \bm{\varphi}^{(l+1)}+\lambda
    \end{bmatrix}^{-1} \notag\\
    &:=
    \begin{bmatrix}
    (\bm{A}^{(l)})^{-1} & \bm{a}^{(l+1)} \\
    (\bm{a}^{(l+1)})^{\mathsf{H}} & a^{(l+1)}
    \end{bmatrix}^{-1}.
\end{align}
Then, by applying the Sherman--Morrison formula, one can obtain
\begin{align}
    &\bm{A}^{(l+1)} = \notag\\
    & \ \ \begin{bmatrix}
      \bm{A}^{(l)} + \bm{A}^{(l)} \bm{a}^{(l+1)} \rho^{-1} (\bm{a}^{(l+1)})^{\mathsf{H}} \bm{A}^{(l)} & - \rho^{-1} \bm{A}^{(l)} \bm{a}^{(l+1)} \\
  -\rho^{-1} (\bm{a}^{(l+1)})^{\mathsf{H}} \bm{A}^{(l)} & \rho^{-1}\\
    \end{bmatrix},
    \label{eq:Aplus}
\end{align}
where 
\begin{align}
  \rho = a^{(l+1)} - (\bm{a}^{(l+1)})^{\mathsf{H}} \bm{A}^{(l)} \bm{a}^{(l+1)}.
\end{align}
By using \eqref{eq:Aplus}, we can sequentially calculate the matrix $\bm{A}^{(l+1)}$ by using $\bm{A}^{(l)}$ with the computational cost of $O(l^2)$. Thus, the total computational cost for Algorithm~\ref{alg:greedy} regarding the computaion of $\bm{A}$ can be reduced to $O(NL^3)$.

\vspace{-4pt}
\subsection{Secondary source placement in broadband case}
\vspace{-4pt}
We defined the cost function $J(\mathcal{S})$ and derived its optimization algorithm for the single frequency case. However, in the sound field synthesis, the target frequency is usually broadband. Our proposed method can be extended to the broadband case by using the weighted sum of the cost function for the $f$th frequency bin, $J_f(\mathcal{S})$, as
\begin{equation}
    J_{\mathcal{F}}(\mathcal{S}) = \sum_{f\in\mathcal{F}} \gamma_f J_f(\mathcal{S}),
\end{equation}
where each $J_f(\mathcal{S})$ is given by \eqref{eq:costfunc} for the $f$th frequency bin, $\gamma_f$ is a positive constant of weighting factor, and $\mathcal{F}$ represents the set of the target frequency bins, $f\in\mathcal{F}$. 
$\gamma_f$ is determined by which frequency is important, e.g, $\gamma_f$ is 1  
The greedy algorithm as in Algorithm~\ref{alg:greedy} and its efficient version can also be similarly derived.

\vspace{-6pt}
\section{Experiments}
\label{sec:4}
\vspace{-6pt}

\begin{figure}[t]
  \centering
  \vspace{-12pt}
  \includegraphics[width=180pt,clip]{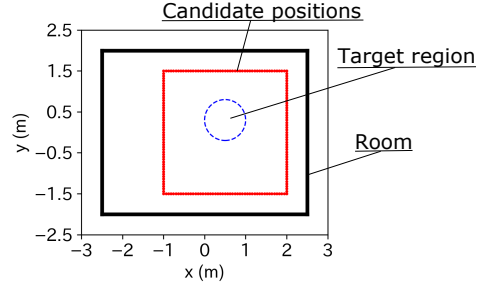}
  \vspace{-7pt}
  \caption{Experimental setup. Blue circular dashed line and red square dots indicate the target region and candidate positions, respectively.}
  \label{fig:field}
  \vspace{-10pt}
\end{figure}

We conducted numerical simulation in a reverberant environment to evaluate the proposed method. The room dimension was $5.0~\mathrm{m} \times 4.0~\mathrm{m}$ and the coordinate origin was set at the center of the room. The reverberant environment was simulated using the image source method~\cite{allen:JASA1979}. The reflection coefficients of the walls were $0.8$. As shown in Fig.~\ref{fig:field}, $200$ candidate source positions were set on the square boundary of $3.0~\mathrm{m} \times 3.0~\mathrm{m}$ size at equal intervals. The target region was a circular region of $0.5~\mathrm{m}$ radius with the center at $(0.5, 0.3)~\mathrm{m}$. We compared the proposed method (Proposed) for narrowband and broadband cases with two regular placements selected from the candidates as far as possible, which will be typically used empirical placements. The number of loudspeaker ($L$) was set to 20.


The desired sound field was a planewave field whose propagation angle was from $-45^\circ$ to $45^\circ$. One of the regular placements (Regular A) was selected from the candidate positions inside the range of the possible incoming planewave angles, which was between intersection points of the square of the candidate positions and two lines of slopes $-45^\circ$ and $45^\circ$ tangent to the circle of the target region. The other regular placement (Regular B) was selected from all the candidate positions. In Proposed, the mean and covariance of $\bm{b}$, $\bm{\mu}$ and $\bm{\Sigma}$ in \eqref{eq:costfunc}, were analytically calculated from the expansion coefficients of the planewave field. 
Each secondary source was assumed to be a point source. The regularization parameter $\lambda$ was $10^{-5}$ in the source selection and $\sigma_{\max}(\bm{C}^{\mathsf{H}}  \bm{W}\bm{C})\times10^{-3}$ in the sound field synthesis, where $\sigma_{\max}(\cdot)$ represents the maximum eigenvalue.
As an evaluation measure, we define the signal-to-distortion ratio (SDR) as
\begin{align}
  \mathrm{SDR}(\omega) = 10\log_{10}\frac{\int_\Omega |u_{\mathrm{des}}(\bm{r},\omega)|^2 \mathrm{d}\bm{r}}{\int_\Omega |u_{\rm des}(\bm{r},\omega)-u_{\mathrm{syn}}(\bm{r},\omega)|^2 \mathrm{d}\bm{r}},
\end{align}
where the regional integral was calculated by discretizing $\Omega$ at intervals of $0.01~\mathrm{m}$. 

\begin{figure}[t!]
    \centering
    \vspace{-18pt}
    \subfloat[Proposed]{\includegraphics[width=120pt,clip]{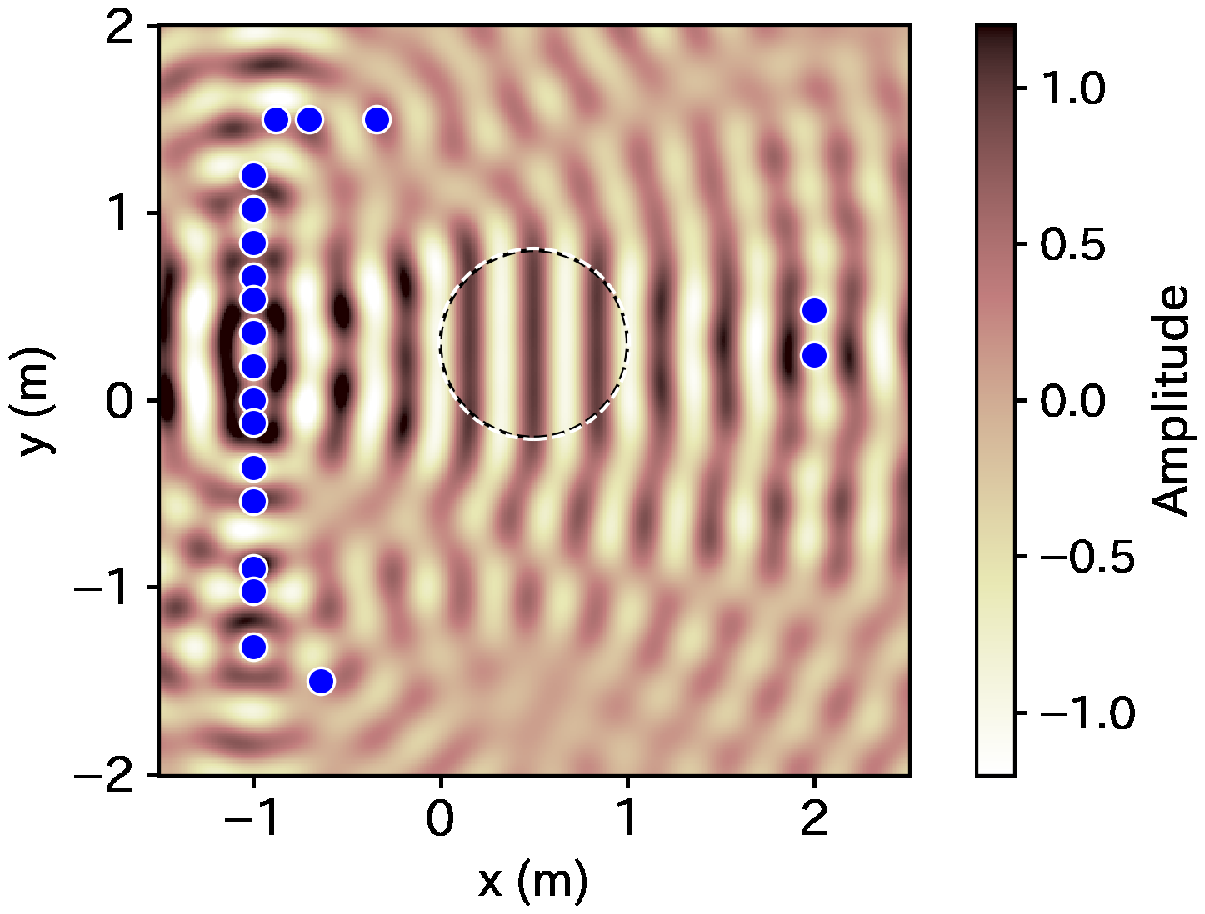}}%
    \subfloat[Regular A]{\includegraphics[width=120pt,clip]{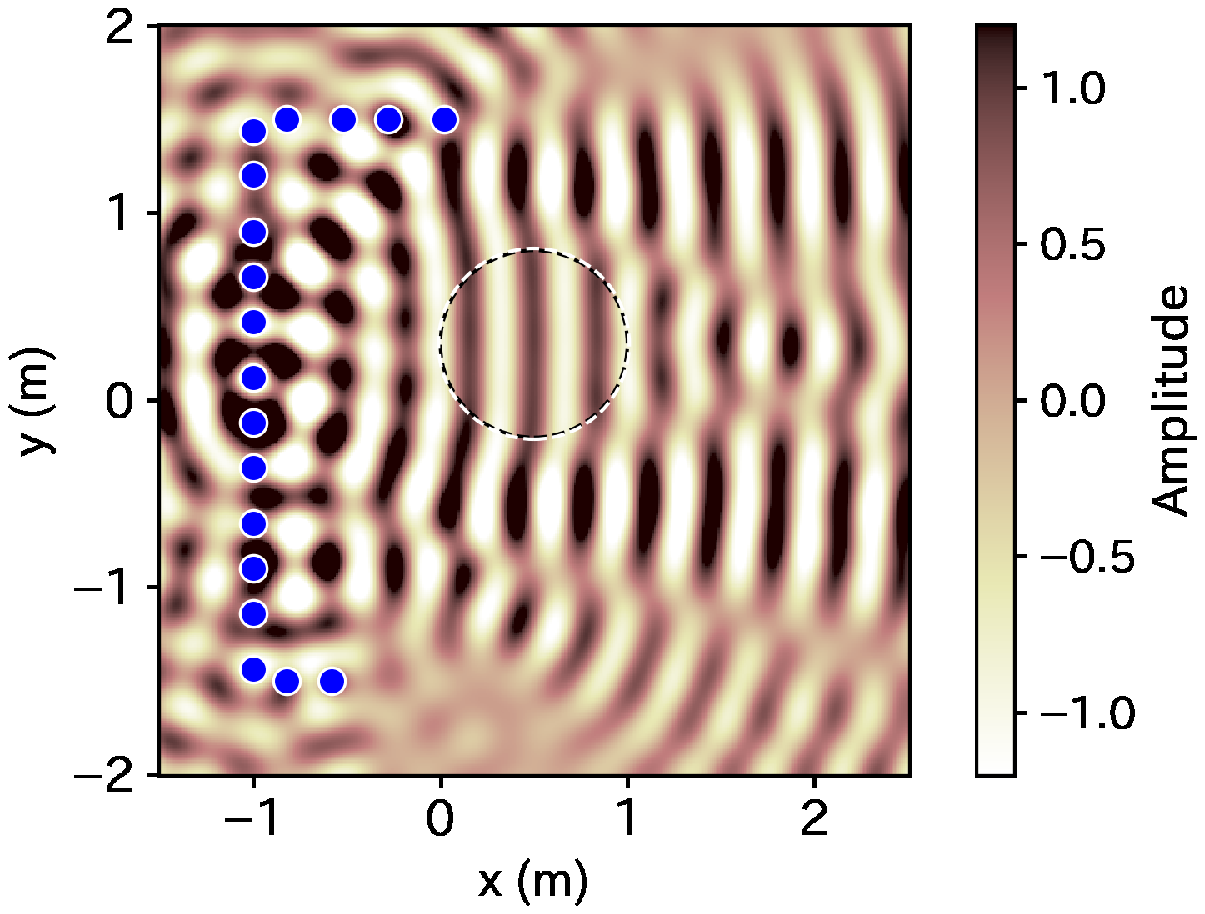}}
    \vspace{-5pt}
    \caption{Pressure distribution of the synthesized sound field at 1000~Hz. The propagation direction of the planewave was $0^{\circ}$. Blue dots indicate the selected loudspeakers.}
    \label{fig:wmmreproduction_p}
    \vspace{-5pt}
    \subfloat[Proposed]{\includegraphics[width=120pt,clip]{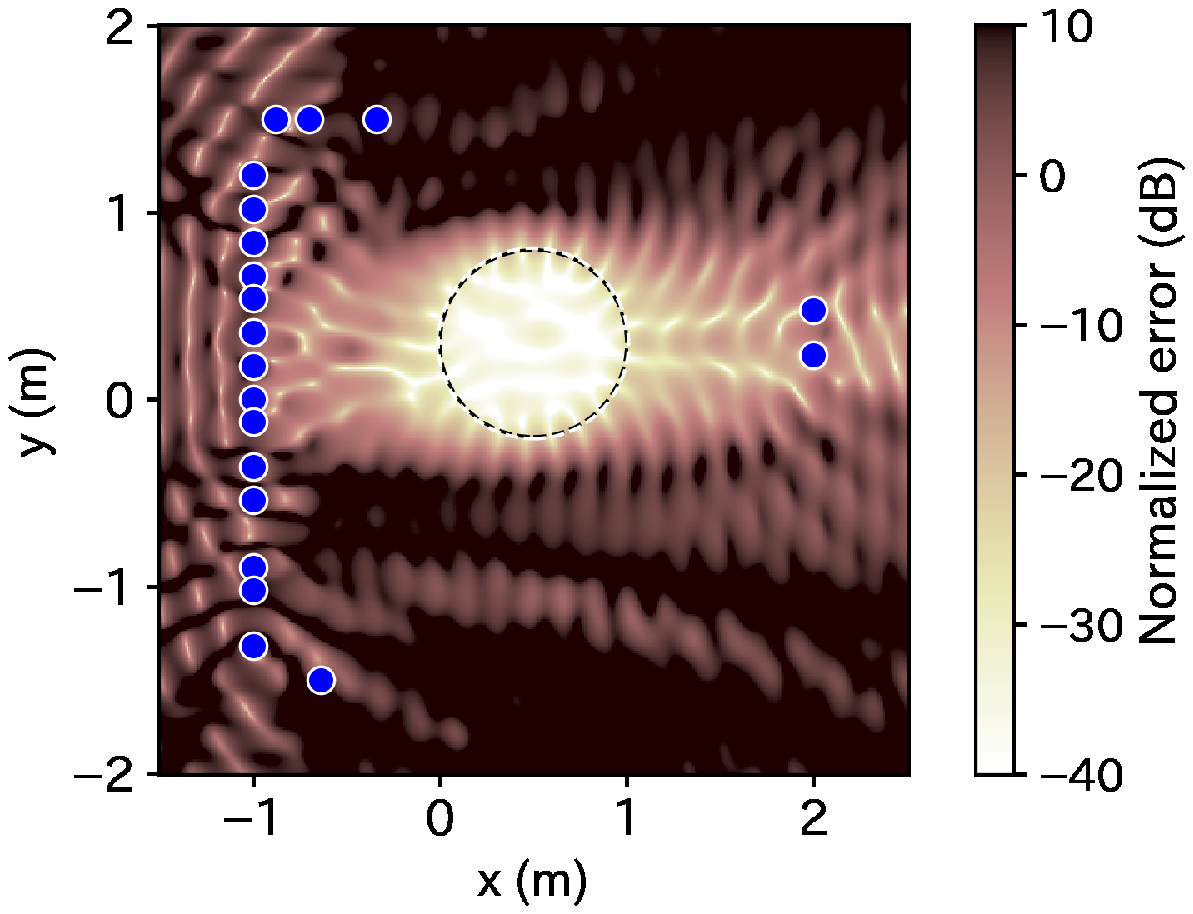}}%
    \subfloat[Regular A]{\includegraphics[width=120pt,clip]{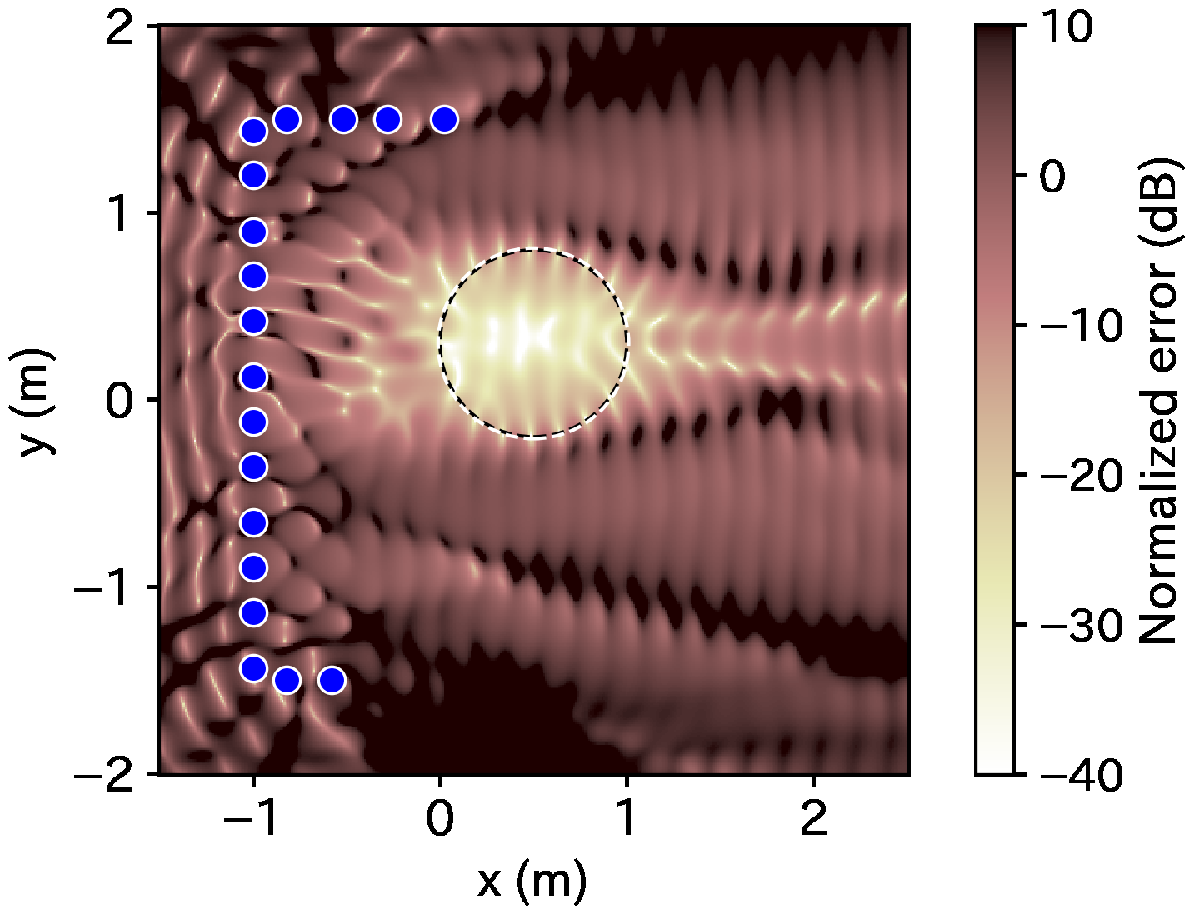}}
    \vspace{-5pt}
    \caption{Normalized error distribution of Fig.~\ref{fig:wmmreproduction_p}. SDRs were 31.5~dB, 20.0~dB, and 18.1~dB in Proposed, Regular A, and Regular B, respectively.}
    \label{fig:wmmreproduction_e}
    \centering
    \vspace{7pt}
    \includegraphics[width=190pt,clip]{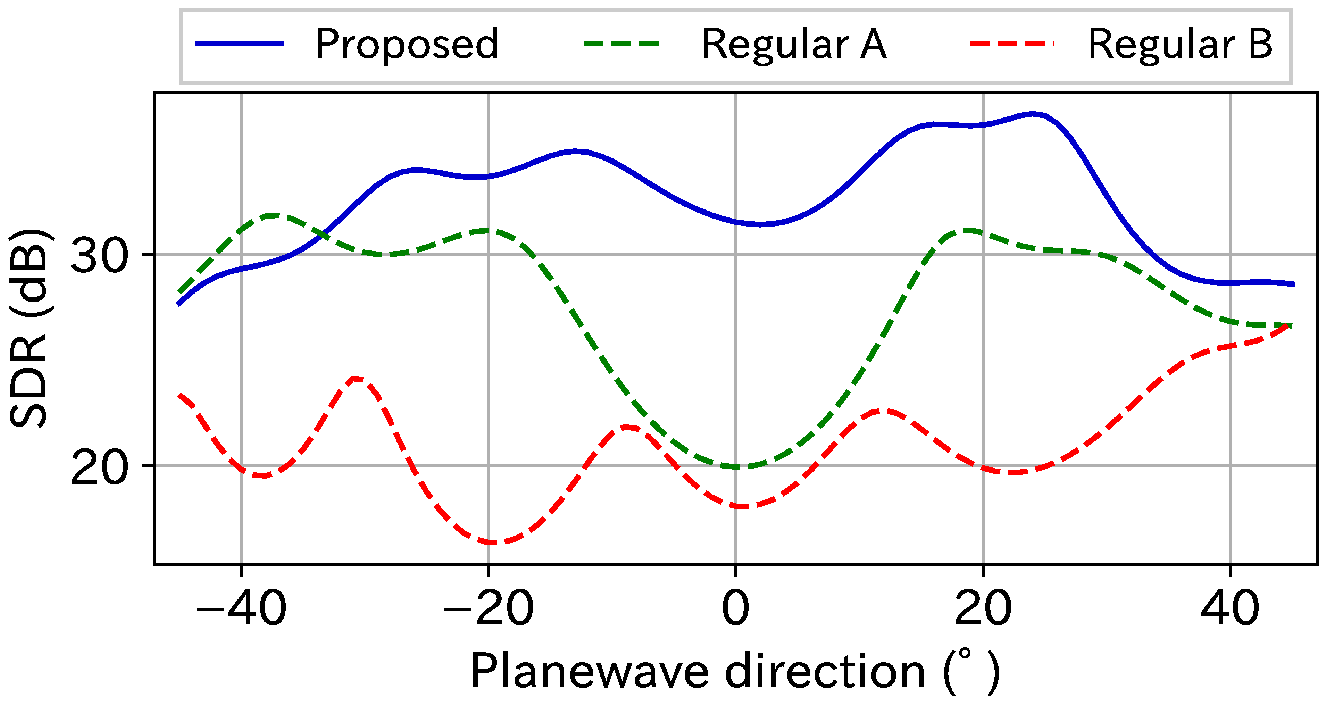}
    \vspace{-5pt}
    \caption{SDR for each propagation direction of planewave at 1000~Hz.}
    \label{fig:angle}
    \vspace{-5pt}
\end{figure}
\begin{figure}[t]
    \centering
    \vspace{-5pt}
    \includegraphics[width=200pt,clip]{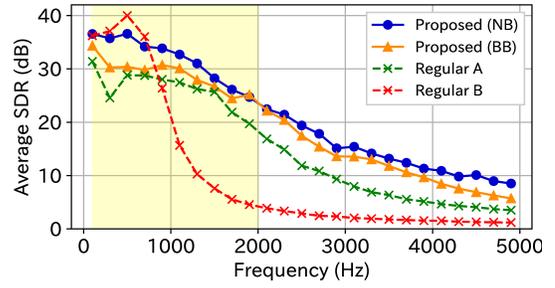}
    \vspace{-5pt}
    \caption{Average SDR with respect to frequency. The proposed placement methods for narrowband (NB) and broadband (BB) cases compared with regular placement methods. The range of target frequencies in broadband case is shown in yellow. }
    \label{fig:fbin}
    \vspace{-10pt}
\end{figure}
Figs.~\ref{fig:wmmreproduction_p} and \ref{fig:wmmreproduction_e} shows the synthesized pressure and normalized error distributions for the propagation angle of $0^\circ$ with the loudspeakers selected by Proposed and Regular A. The frequency of the planewave was 1000~Hz.
The loudspeaker placement of Proposed was more concentrated around the left side of the target region, compared with Regular A. Besides, two loudspeakers were placed on the right side in Proposed. 
Fig.~\ref{fig:angle} shows the SDR when synthesizing desired planewaves of each angle. 
The SDRs of Proposed was uniformly high for all the directions whereas those of Regular A was dependent on the planewave angles. The SDRs of Regular B were lower than the others for all the directions. 
The average SDR of Proposed, Regular A, and Regular B were 32.5, 27.6, and 21.0~dB, respectively. 

We also performed an evaluation in the broadband case. The target frequency bins were set from 100 to 2000~Hz at intervals of 100~Hz. The weighting factor $\gamma_f$ was 1 for all the frequency bins. We also evaluated the proposed placement method for the narrowband case, where the placement of 20 loudspeakers was optimized at each frequency.
Fig.~\ref{fig:fbin} shows the average SDR with respect to frequency. The proposed placement methods for narrowband and broadband cases were denoted as Proposed (NB) and Proposed (BB), respectively. Both the proposed methods consistently outperformed Regular A. In particular, Proposed (BB) achieved high reproduction accuracy even at above the target frequencies, compared with Regular A. Although the average SDR of Regular B was high at low frequencies, it significantly deteriorated above 1000~Hz. 

\vspace{-6pt}
\section{Conclusion}
\label{sec:5}
\vspace{-6pt}

We proposed a mean-square-error-based secondary source placement method in sound field synthesis. We formulated a cost function of the source placement for the weighted mode matching, but it can be applied to general linear-least-squares-based sound field synthesis methods by replacing the variables of transfer functions and desired field. The proposed cost function is based on the mean square error of the reproduced sound field, incorporating the statistical properties of possible desired sound fields. An efficient greedy algorithm for minimizing the proposed cost function is also derived for the case that the target frequencies are the narrowband and broadband. In numerical experiments, a high reproduction accuracy was achieved by the proposed source placement compared with the empirical regular placement.

\vspace{-6pt}
\section{Acknowledgment} 
\label{sec:ack}
\vspace{-6pt}

This work was supported by JST PRESTO Grant Number JPMJPR18J4, and JSPS KAKENHI Grant Number
JP19H01116.

\vfill\pagebreak

\label{sec:refs}

\bibliographystyle{IEEEtran}
\bibliography{str_def_abrv,refs21,refs_sk,koyama_en}

%
%
%
%
%
%
%
%
%

\end{sloppy}
\end{document}